\documentclass[sort&compress,APS,STIX2COL]{WileyNJDv5} %STIX1COL,STIX2COL,STIXSMALL AMA,Times2COL
\usepackage{colortbl}
\usepackage{wrapfig}

\usepackage{textcomp}

\usepackage{siunitx}[=v2]
\DeclareSIUnit\fps{fps}

\usepackage{csquotes}

\usepackage{nicefrac}

\usepackage[
	nonumberlist, 
	acronym, 
	nomain,
	nopostdot,
]{glossaries}

\newacronym{isru}{ISRU}{\textit{in-situ} resource utilisation}
\newacronym{gtb}{GTB}{GraviTower Bremen Pro} 
\newacronym{zarm}{ZARM}{Center of Applied Space Technology and Microgravity}
\newacronym{1g}{$1g$}{Earth gravity}
\newacronym{mug}{$\mu g$}{microgravity}
\newacronym{dem}{DEM}{discrete element methods}
\newacronym{iss}{ISS}{international space station}
\newacronym{afm}{AFM}{atomic force microscopy}
\newacronym{vdw}{vdW}{van der Waals}
\newacronym{lro}{LRO}{Lunar Reconnaissance Orbiter}
\definecolor{brickred}{rgb}{0.8, 0.25, 0.33}
\definecolor{darkorange}{rgb}{1.0, 0.55, 0.0}
\definecolor{persiangreen}{rgb}{0.0, 0.65, 0.58}
\definecolor{persianindigo}{rgb}{0.2, 0.07, 0.48}
\definecolor{cadet}{rgb}{0.33, 0.41, 0.47}
\definecolor{turquoisegreen}{rgb}{0.63, 0.84, 0.71}
\definecolor{sandybrown}{rgb}{0.96, 0.64, 0.38}
\definecolor{blueblue}{rgb}{0.0, 0.2, 0.6}
\definecolor{ballblue}{rgb}{0.13, 0.67, 0.8}
\definecolor{greengreen}{rgb}{0.0, 0.5, 0.0}
\definecolor{razzmatazz}{rgb}{0.89, 0.15, 0.42}
\definecolor{ultramarine}{rgb}{0.07, 0.04, 0.56}
\definecolor{midnightgreen}{rgb}{0.0, 0.29, 0.33}
\definecolor{lavenderpurple}{rgb}{0.59, 0.48, 0.71}
\definecolor{bittersweet}{rgb}{1.0, 0.44, 0.37}
\definecolor{amaranth}{rgb}{0.9, 0.17, 0.31}
\definecolor{patriarch}{rgb}{0.5, 0.0, 0.5}
\definecolor{darkcandyapplered}{rgb}{0.64, 0.0, 0.0}
\definecolor{cadmiumgreen}{rgb}{0.0, 0.42, 0.24}
\definecolor{darkgreen}{rgb}{0.0, 0.2, 0.13}
\definecolor{deepcarrotorange}{rgb}{0.91, 0.41, 0.17}
\definecolor{deepcarmine}{rgb}{0.66, 0.13, 0.24}
\definecolor{maroon}{rgb}{0.69, 0.19, 0.38}
\definecolor{midnightblue}{rgb}{0.1, 0.1, 0.44}
\definecolor{lava}{rgb}{0.81, 0.06, 0.13}
\definecolor{ceruleanblue}{rgb}{0.16, 0.32, 0.75}
\definecolor{sacramentostategreen}{rgb}{0.0, 0.34, 0.25}
\definecolor{darkgreen}{rgb}{0.12, 0.3, 0.17}%
\definecolor{carmine}{rgb}{0.92, 0.3, 0.26}%

\articletype{Article Type: Invited Contribution}%

\journal{Journal for Numerical and Analytical Methods in Geomechanics}
\volume{00}
\copyyear{2025}
\begin{document}

\title{An Open Database of Lunar Regolith and Simulants Properties}

\author[1]{L\'{e}onie Gasteiner}
\author[1]{Naomi Murdoch}
\author[1,2]{Olfa D'Angelo}

\authormark{Gasteiner \textsc{et al.}}
\titlemark{An Open Database of Lunar Regolith and Simulants Properties}

\address[1]{\orgdiv{Institut Sup\'{e}rieur de l'A\'{e}ronautique et de l'Espace (ISAE-SUPAERO)}, \orgname{Universit\'{e} de Toulouse}, \orgaddress{10 Avenue Marc P\'{e}legrin, 31055 Toulouse}, \country{France}}

\address[2]{\orgdiv{Van der Waals-Zeeman Institute, Institute of Physics}, \orgname{University of Amsterdam}, \orgaddress{Science Park 904, Amsterdam 1098XH}, \country{The Netherlands}}

\corres{Corresponding author: Olfa D'Angelo.\email{o.dangelo@uva.nl}}

%\presentaddress{This is sample for present address text this is sample for present address text.}

%\fundingInfo{Text}
%\JELinfo{ejlje}
%by the Apollo, Luna, and Chang'E programs, 

\abstract[Abstract]{
Lunar regolith, the layer of unconsolidated material covering the Moon's surface, is central to the science and technology developed for the Moon,
notably related to in-situ science investigations, resource utilization, surface infrastructure, and mobility systems.
However, data on lunar soil properties remain fragmented across decades of mission reports, often in formats that are difficult to access or interpret.
We present a newly compiled database of lunar regolith physical and geotechnical properties,
including data collected by 
direct in-situ measurements from crewed missions, 
estimates inferred from surface interactions on the Moon and using remote sensing,
as well as laboratory analyses of samples returned to Earth.
The data collected include, among others, the angle of internal friction and cohesion (both Mohr-Coulomb model parameters), 
bulk density, and static bearing capacity,
extracted from Luna and Apollo-era historical mission documentation
all the way to contemporary Lunar programs.
The dataset specifies the type and location of the tests from which each value was obtained.
Our database also includes parameters for lunar regolith simulants, providing a direct link between mission data and laboratory studies.
In addition to centralizing this information, we developed a user interface that facilitates data retrieval, filtering, and visualization. 
This interface enables users to generate customized plots for comparative analysis.
Developed in an open-science perspective, it is designed to evolve in response to the community's needs.
The database and its associated tools significantly enhance the accessibility and usability of lunar regolith and simulants data for scientific and engineering research.
}


\jnlcitation{\cname{%
\author{Gasteiner L.},
\author{Murdoch N.}, and
\author{D'Angelo O.}}.
\ctitle{An Open Database of Lunar Regolith and Simulants Properties} \cjournal{\it J Comput Phys.} \cvol{2026;00(00):1--18}.}

\maketitle

\renewcommand\thefootnote{}
%\footnotetext{\textbf{Abbreviations:} ANA, anti-nuclear antibodies; APC, antigen-presenting cells; IRF, interferon regulatory factor.}

\renewcommand\thefootnote{\fnsymbol{footnote}}
\setcounter{footnote}{1}

\section{Introduction}\label{sec:intro}

The physical and geotechnical properties of lunar regolith govern a wide range of processes, from rover mobility to landing site selection, excavation strategies, instrument design and construction methods on the Moon \cite{toklu_construc_2022}.
With the renewed interest in lunar exploration,
driven by NASA's Artemis program \cite{noble_artemis_2024}, the emergence of commercial actors \cite{LIN_return_moon}, and the growing number of nations pursuing independent lunar missions \cite{Flahaut2023, chandrayaan3_2025, change5_li_2022,change3_Dong_2017,change4_tang_2020},
a new generation of scientists and engineers is turning to existing data on the lunar surface.

For these users, there is a growing need for data that are accessible, comparable, and interpretable.
Because of the high ecological and economic costs of space missions, 
all available data are highly valuable.
Much of our current understanding stems from the Apollo-era missions, notably compiled in the Lunar Sourcebook \cite{LunarSourcebook1991}.
More recent missions \cite{change3_Dong_2017,change5_li_2022,change4_tang_2020,chandrayaan3_2025} but also re-interpretation of existing data \cite{gromov_physical_1998,slyuta_physical_2014,LunarSourcebook1991}
are now expanding this foundation.
Yet, many critical parameters remain difficult to retrieve. 
This is largely due to the decentralized and heterogeneous nature of historical lunar mission data.
Typically, reports from the Apollo, Luna, and Surveyor programs
are often embedded in incompatible formats, dispersed across archives, or accessible only through scanned documents in their original language, which complicates their integration, comparison, and analysis. 
This fragmentation hinders both the advancement of lunar science and the reinterpretation of existing results using modern analytical tools.

We present a newly compiled database of lunar regolith physical and geotechnical properties,
including data collected by 
direct in-situ measurements from crewed missions, 
estimates inferred from surface interactions on the Moon and using remote sensing,
as well as laboratory analyses of samples returned to Earth.
In line with recent efforts to revisit Apollo-era datasets and reconstruct a consistent picture of lunar soil mechanics \cite{Connolly2023, Seifamiri2024},
our work integrates data ranging from 1966 with the Luna 9 mission report \cite{luna_9_1966}, to 2025 with a paper on the Chandrayaan-3 mission \cite{chandrayaan3_2025}.
%covering missions from Surveyor, Apollo, and Luna to the latest Chang'E programs.
The database includes results from direct in-situ testing during crewed missions \cite{hess_apollo11_1969,stephenson_apollo12_1970,herbert_apollo14_1971,calio_apollo15_1971,hinners_apollo16_1972,hinners_apollo17_1973}; 
estimations derived from surface interactions such as rover tracks \cite{cherkasov_soviet_1973,herbert_apollo14_1971,calio_apollo15_1971,hinners_apollo17_1973,tang_change4_2020}, astronaut footprints \cite{hinners_apollo16_1972}, lander padprints \cite{jaffe_surveyorI_1966,haglund_surveyorIII_1967,jaffe_surveyorV_1967,jaffe_surveyorVI_1968,hess_apollo11_1969,stephenson_apollo12_1970};
and remote sensing data \cite{bickel_LRO_boulder_tracks_2019,bickel_boulder_tracks_crewrover_2020}.

Data on lunar regolith simulants are also included \cite{maki_FJS-1_2014, alshibli_JSC-1A_2009, anbazhagan_LSS-ISAC-1_2021, chancharoen_TLS-01_2021, engelschion_eac-1a_2020, Gasteiner2025, he_chunmei_GRC-3_2013, he_CUG-1A_2010, zou_polyU-1_2024, zheng_cas-1_2009, yin_LHS-1LMS-1_2023, teng_WHU-1_2025, tang_CLDS-1_2017, stoeser_BP-1_nodate, ryu_KLS-1_2018, ruan_IGG-01_2024, noauthor_OB-1_nodate-1, noauthor_JSC-1A_nodate, li_nao1_2009}.
Such materials are essential for testing technologies and experimental methods intended for lunar applications. 
However, their mechanical and rheological properties often differ significantly from those of true lunar regolith, especially from true lunar regolith as found \emph{on the Moon}. 
The development of improved simulants requires a robust ground-truth reference 
-- or rather, a \textit{Moon-truth}.

Previous efforts to centralize lunar simulant properties, notably the Colorado School of Mines Planetary Simulant Database \cite{Laughery2025_SimulantsDatabase},
provided valuable benchmarks for simulant properties.
Recent disruptions in access to such repositories have further highlighted the need for a permanent, open-access platform that integrates historical mission data with contemporary simulant research.

In addition to the geotechnical results themselves, 
our database specifies the type of test used to measure each parameter (e.g. penetrometer, rover tracks analysis, direct shear test, etc.).
In an effort towards greater reproducibility, 
when available, complementary details, 
such as the unconsolidated bulk density or the range of normal stresses applied in the Mohr-Coulomb analysis, are included.
When relevant and available, it also includes contextual information, 
such as the geographical coordinates of in-situ tests, local gravitational acceleration, and environmental conditions during experiments. 
This information is particularly relevant given the increasing recognition that reduced gravity,
among other aspects of the lunar environment, 
significantly influence granular behavior \cite{wilkinson_granular_2005,Murdoch2013,Chaikin2020,DAngelo2022,Sanchez2023,Madden2025,Gaida2025,DAngelo2025}.

This article is organized as follows:
Section~\ref{sec:quickstart} proposes a guide to quickly start using the database.
Section~\ref{sec:methodology} describes the structure and methodology underlying it, 
including data selection and formatting choices. 
Section~\ref{sec:usecases} presents three example use-cases that illustrate how the database can support current and future research. 
Finally, in Section~\ref{sec:discussion}, we discuss the limitations of the current version and outline directions for future development, emphasizing the importance of community engagement in fostering an open-science framework for lunar geotechnical research.

\section{Quick start}\label{sec:quickstart}

The Lunar Regolith Database is openly accessible via a web application hosted on the Streamlit platform at \url{https://lunar-regolith-database.streamlit.app/} or on the hugging face platform at \url{https://huggingface.co/spaces/leoniegasteiner/Lunar-Regolith-Database}. Access requires only an internet connection and a web browser.

For users who wish to run the database locally, the full repository containing both the source code and the dataset can be downloaded from \url{https://github.com/leoniegasteiner/Lunar-Regolith-Database}. Detailed installation and usage instructions are provided in the accompanying \texttt{README.md} file. A concise overview is given below.

\subsection{Requirements}
To run locally, the database requires Python (version 3.9–3.12) and the following packages: \texttt{streamlit}, \texttt{pandas}, \texttt{numpy}, \texttt{plotly}, \texttt{re}, \texttt{base64}, \texttt{importlib}, and \texttt{os}. The application runs in any modern browser (Chrome, Firefox, Safari, or Edge) and does not require an internet connection once installed locally.

\subsection{Execution}
To launch the application, navigate to the project directory in a terminal and execute the following command:
\begin{verbatim}
streamlit run Lunar_Database.py
\end{verbatim}
The application will automatically open in the default browser. Alternatively, it can be accessed manually through the local URL provided in the terminal output.

\section{Methodology}\label{sec:methodology}

\subsection{Data sources and extraction}

We systematically reviewed technical documents from multiple lunar missions, prioritizing those that reported in-situ measurements and laboratory analyses of returned samples. 
Key sources included:
\begin{itemize}
\item NASA Surveyor (1966 - 1968) preliminary science reports \cite{jaffe_surveyorI_1966,haglund_surveyorIII_1967,jaffe_surveyorV_1967,jaffe_surveyorVI_1968,jaffe_surveyorVII_1968-1};
\item NASA Apollo mission (1969 - 1972) technical reports \cite{hess_apollo11_1969,stephenson_apollo12_1970,herbert_apollo14_1971,calio_apollo15_1971,hinners_apollo16_1972,hinners_apollo17_1973}, and sample catalogues \cite{A11_sample_catalogue_1977,A12_sample_catalogue_1970,A14_sample_catalogue_1978,A15_sample_catalogue_1971,A16_sample_catalogue_1972,A17_sample_catalogue_1973};
\item Soviet Luna mission (1966 - 1976) summaries \cite{cherkasov_soviet_1973, slyuta_physical_2014, slyuta_physical_2014,luna_9_1966};
\item Recent mission publications, including data from the Chang'E and Chandrayaan programs \cite{dong_Change3_2017,change4_tang_2020,change5_li_2022,chandrayaan3_2025}.
\item Later interpretations of lunar data \cite{LunarSourcebook1991,gromov_physical_1998,slyuta_physical_2014,sample_PSD_catalog}
\end{itemize}

Data extraction was performed manually to ensure accuracy, given the often qualitative and sometimes unstructured nature of original sources. 
For in-situ lunar data, each entry in the database includes metadata specifying the mission, testing method, geographic location (latitude and longitude on the Moon), and the corresponding reference source.
The \enquote{location} field provides the lunar coordinates of the mission landing. 
When specified in the source, we include the model used to derive parameters such as cohesion or bearing capacity in the comments column of the specific entry line along with other assumptions made for the measurement.
For data obtained on Earth from returned lunar samples, 
the database records the mission responsible for sample return, and the location of the landing on the moon.
A clear distinction between in-situ surface measurements and post-mission, ground-based analyses is maintained through the \enquote{test location} field, distinct from \enquote{location}:
the test location is either \enquote{in-situ}, \enquote{on Earth}, or \enquote{remote} for remote sensing missions data.

For regolith simulants, the database includes the developer, year of creation, simulant type (mare or highland), and the corresponding mechanical parameters derived from the lunar regolith when available.  
A dedicated page within the database, accessible via a button in the main filter panel, provides access to simulants data through a searchable table and several visualization options.  
To facilitate direct comparison between simulants and actual lunar regolith, both datasets report the same main physical parameters.

A comprehensive table combining lunar regolith and simulant data is also available for download from the main interface.  
Users can apply filters to tailor the displayed data to their specific research needs.

A page containing information about lunar samples brought back by the Apollo and Luna missions is accessible through the sidebar. 
This section includes a table with general information and description of the lunar samples including the sample ID number, weight, and a physical description of the rocks and soils. 
A table containing the particle size distribution of soil samples is also available, with an interactive plotting section allowing for comparison between samples of different missions and collection depths.

A page displaying details about each specific missions context and data can also be accessed through the main filter panel. It elaborates on the context of each mission's test and sample analysis. All the data collected during that specific mission is displayed in a table, and another table displays information on the samples brought back from that mission.

%General interpretations of lunar regolith are often included in the different mission reports and reinterpretation papers, these values corresponding to general parameters for bulk lunar regolith are included in a different page in the database.

The physical parameters included in the database were extracted from multiple sources, beginning with the earliest available reports and incorporating subsequent interpretations when possible.  
When several values were reported for the same parameter, mission, and testing method, all were retained to provide a comprehensive overview of the data.  
The year of publication was also recorded to help users assess the evolution of interpretations over time.

\subsection{Parameters}
To accurately reflect geomechanical testing principles, the database categorizes parameters into metadata, direct field measurements, laboratory testing conditions, and derived geomechanical properties. 
The current version of the database includes (but is not limited to):

\subsubsection*{Metadata}
\begin{itemize}
    \item \textbf{Mission, Location, Terrain, and Year:} parameters specific to the mission considered, with the location of the spacecraft landing, the geologic terrain of the region explored, and the year of the mission;
    \item \textbf{Test:} the measurement method used, currently among the ones described in Sec.\,\ref{subsec:TestingMethods};
    \item \textbf{Test location:} indicates whether the measurement was performed in-situ on the lunar surface, on Earth using returned samples, or remotely for remote sensing data;
    \item \textbf{Testing environment:} specifies the atmosphere in which the test was performed; in vacuum for in-situ and remote measurements, and in vacuum, nitrogen, or air for returned samples;
    \item \textbf{Sample ID:} specifies the identification number of the sample on which the test was performed, when applicable;
    \item \textbf{Reference and Comments:} bibliographic citation of the source from which the data was extracted and additional information relevant to the specific entry line. 
\end{itemize}

\vspace{-0.8cm}

\subsubsection*{Direct Field Measurements}
\begin{itemize}
    \item \textbf{Depth (\si{\centi\meter}):} depth below the lunar surface at which the measurement was taken, when available;
    \item \textbf{Force applied (\si{\newton}):} for penetrometer tests, the force applied on the soil during the test, when available;
    \item \textbf{Static bearing pressure (\si{\kilo\pascal}):} the estimated pressure exerted on the regolith by lander footpads, astronaut footprints, or rover wheels based on their respective weights in lunar gravity;
    \item \textbf{Contact area (\si{\square\centi\meter}):} area over which the applied force or static bearing pressure is applied.
\end{itemize}

\vspace{-0.8cm}

\subsubsection*{Laboratory Testing Conditions}
\begin{itemize}
    \item \textbf{Normal stress range (\si{\kilo\pascal}):} range of normal stresses applied during shear testing of returned samples on earth from which the Mohr-Coulomb parameters were derived;
\end{itemize}

\vspace{-0.8cm}

\subsubsection*{Derived and Inferred Mechanical Properties}
\begin{itemize}
    \item \textbf{Angle of internal friction (\si{\degree}):} first parameter in the Mohr-Coulomb model, corresponding to the slope of the Coulomb failure line \cite{labuz_mohrcoulomb_2012};
    \item \textbf{Cohesion (\si{\kilo\pascal}):} second parameter in the Mohr-Coulomb model, corresponding to the intercept of the Coulomb failure line with the shear strength axis \cite{labuz_mohrcoulomb_2012};
    \item \textbf{Bulk density (\si{\gram\per\cubic\centi\meter}):} defined as the mass of unconsolidated regolith per unit volume;
    \item \textbf{Bearing capacity (\si{\kilo\pascal}):} defined as the maximum load per unit area that the regolith can support without failure;
    \item \textbf{Cone penetration resistance gradient (\si{\kilo\newton\per\square\meter\per\meter}):} the rate at which the soil's resistance to penetration increases with depth, calculated from the applied force and penetrometer geometry;
    \item \textbf{Porosity (\%) and Void ratio:} interconvertible expressions of the volume of voids relative to the total or solid volume of the soil;
    \item \textbf{Density of grains (\si{\gram\per\cubic\centi\meter}):} the mass of the solid soil particles per unit volume of solids;
    \item \textbf{Compressibility coefficient:} the soil's volume change behavior under applied normal stress.
\end{itemize}

The database will continue to be updated over time, reflecting new findings and contributions. In line with our commitment to open science, we actively invite researchers to contribute their own data, particularly on existing or novel lunar regolith simulants.
To ensure the scientific integrity and reproducibility of the database, all submitted data must meet the following criteria:
The data must be published in a peer-reviewed, non-predatory journal.
The testing methodology must be clearly described. If no detailed protocol is available in the publication or online, contributors are required to provide a comprehensive description of the procedure.
For all datasets currently included in the database, the associated testing methods are described in detail in the sections below.

\subsection{Testing methods classification}\label{subsec:TestingMethods}

The database categorizes the data obtained on the Moon according to the testing methodology used. Due to the limited number of experiments that could be performed in-situ, no two investigation methods were identical. 
The data were therefore divided according to the general procedure used to obtain them. 
The paragraphs below describe the terminology used for the classification of these testing methods, the general procedure they represent, and the missions in which they were applied. 
The different testing methods are classified depending on the testing location; in-situ, on Earth, or remote sensing.

\subsubsection{in-situ testing methods}

\begin{description}
\item 
\textbf{Spacecraft Touchdown Analysis \& Vernier Thrusters:}
Spacecraft touchdown analysis was the first method used to retrieve data on the mechanical properties of the lunar regolith. 
It was applied during most lunar missions; however, numerical values were only determined for the Surveyor missions I, III, V, and VI, as well as the Apollo 11 and 12 missions \cite{jaffe_surveyorI_1966,haglund_surveyorIII_1967,jaffe_surveyorV_1967,jaffe_surveyorVI_1968,hess_apollo11_1969,stephenson_apollo12_1970}. 
The method consists of analysing the appearance of the lunar surface near the lander footpads through photographs taken after disturbance, together with the histories of axial loads recorded in the shock absorbers of each leg during landing. 
Similarly, the Vernier thrusters analysis was performed during the Surveyor VII mission \cite{jaffe_surveyorVII_1968-1}, where the lander was equipped with vernier thrusters used for fine control during descent and landing. 
The analysis of the interaction between the thruster plumes and the lunar surface provided additional information on the regolith's mechanical properties.
This testing methodology required assumptions on the bulk density and cohesion of the regolith to be made to extract values of parameters such as the bearing capacity or the angle of internal friction. \bigskip

\item \textbf{Trench Experiment:}
Trench tests were performed using the lunar sampler of the Surveyor VII mission and by the astronauts of the Apollo 14 and 15 missions \cite{jaffe_surveyorVII_1968-1,herbert_apollo14_1971,calio_apollo15_1971}. 
The test consisted of scooping portions of lunar regolith and observing the angle of repose and the general behavior of the material when digging into it. 
These tests facilitated the derivation of Mohr-Coulomb parameters (cohesion and friction angle), often making assumptions on the density based on previous mission data.

\bigskip
\item 
\textbf{Optical Assessment:}
The optical assessment method was primarily used during the Luna 9 mission \cite{luna_9_1966}, which involved taking photographs of the lunar surface. 
No numerical data on the mechanical properties of the regolith were retrieved using this method. 

\bigskip
\item 
\textbf{Penetrometer:}
Penetrometer experiments were first performed during the Luna 13 mission \cite{cherkasov_soviet_1973}, whose lander was equipped with a mechanical punch penetrometer that provided numerical data on the top layer of the lunar surface. 
Subsequent penetrometer experiments were conducted by the Soviet rover Lunokhod I, deployed by the Luna 17 mission \cite{slyuta_physical_2014}, 
allowing data collection in regions with undisturbed surface soil. 
Similar experiments were also performed by the astronauts of the Apollo 14, 15, and 16 missions \cite{herbert_apollo14_1971,calio_apollo15_1971,hinners_apollo16_1972}. 
These experiments consisted of recording the force required to insert a device into the regolith to a given depth. 
This experiment yielded values for the bulk density, cohesion, angle of internal friction, and in some cases on the void ratio of the lunar soil. 
When available, the force applied to insert the penetrometer in the lunar surface, the cone penetration resistance of the penetrometer tip and its area are reported.

\bigskip
\item 
\textbf{Rover Tracks \& Footprint Analysis:}
The observation of rover tracks provides valuable insight into the behavior of lunar regolith. 
Such observations began with the Lunokhod I and II rovers from the Luna 17 and 21 missions \cite{cherkasov_soviet_1973,slyuta_physical_2014} and continued with the Apollo 14, 15, and 17 missions \cite{herbert_apollo14_1971,calio_apollo15_1971,hinners_apollo17_1973}, as well as more recent missions such as Chang’e 5 \cite{change5_li_2022}. 
Although the Apollo 14 mission did not include a rover, similar observations were made using the tracks of the Modular Equipment Transporter (MET). 
This technique involves focusing cameras on the rover wheels to observe the interaction between the wheels and the lunar surface. 
It allows estimation of regolith density, porosity, and internal friction angle, although cohesion values are often assumed from prior experiments. 
Footprint analysis was performed during the Apollo 11, 12, 14, and 15 missions \cite{hess_apollo11_1969,stephenson_apollo12_1970,herbert_apollo14_1971,calio_apollo15_1971}. 
This method involves analyzing the depth and shape of astronaut footprints on the lunar surface to infer mechanical properties of the regolith.

\bigskip
\item 
\textbf{Drive Tube \& Drill Stem:}
Drive tubes and drill stems were used during the Luna 16, 20, and 24 missions \cite{cherkasov_soviet_1973,slyuta_physical_2014,Druzhininskaya1971}, and in all the Apollo missions \cite{hess_apollo11_1969,stephenson_apollo12_1970,herbert_apollo14_1971,calio_apollo15_1971,hinners_apollo16_1972,hinners_apollo17_1973}. These methods involve extracting cylindrical samples of lunar regolith using various drilling techniques. The samples are then analysed to determine properties such as bulk density and cohesion.
Although these methods primarily served for sample collection, they also provided indirect information on the mechanical properties of the regolith through the behavior of the regolith during and after sampling.

\bigskip
\item 
\textbf{Other Testing Methods:}
With more recent lunar missions, new technologies have enabled the design of dedicated experiments. 
For the sake of simplicity and due to the limited availability of recent data, these methods are not detailed in this paper. 
Notable examples include the Lunar Penetration Radar of the Chang’e 3 mission \cite{dong_Change3_2017} and the Surface Thermophysical Experiment of the Chandrayaan-3 mission \cite{chandrayaan3_2025}, 
which are documented in the references cited in the database. 

\end{description}

\subsubsection{On-Earth testing methods}

\begin{description}
\item 
\textbf{Drive Tube \& Drill Core:}
Drive tube and drill cores brought back to Earth by the Apollo and Luna missions were tested using various techniques, mainly performing petrological and biological examinations. 
In many cases, the analysis of core samples on Earth allowed a precise determination of the regolith's bulk density and particle size distribution and, in some cases, the void ratio and the density of grains.

\bigskip
\item 
\textbf{Direct Shear, Triaxial Shear, \& Rotational Shear:}
These testing methods are standard geotechnical laboratory tests used to determine the shear strength parameters of soils. 
The direct shear test involves applying a horizontal force to a soil sample until failure occurs, allowing for the determination of cohesion and internal friction angle often using the Mohr-Coulomb failure framework.
The triaxial shear test subjects a cylindrical soil sample to axial and confining pressures, providing a more comprehensive understanding of soil behavior under different stress conditions. 
The rotational shear test involves applying a torque to a soil sample to measure its shear strength. 
When available, the range of normal stresses applied to the samples and the compressibility coefficient are specified 

\bigskip
\item 
\textbf{Rotating Drum:}
Rotating drum tests were conducted on samples brought back by the Luna 16 and 20 missions \cite{cherkasov_soviet_1973}. 
They consist of observing the behavior of regolith when placed in a rotating cylinder, 
allowing for the determination of the material's dynamic angle of repose and flowability. % and flowability.
\end{description}

\subsubsection{Remote Sensing testing methods}

\begin{description}
\item 
\textbf{Boulder Tracks:}
Boulder track analysis was performed using images captured by lunar orbiters, such as the \gls{lro}~\cite{bickel_LRO_boulder_tracks_2019,bickel_boulder_tracks_crewrover_2020}. 
This method involves studying the tracks left by boulders rolling down slopes to infer the mechanical properties of the regolith, such as the internal friction angle. 
They often require assumptions based on prior in-situ measurements to estimate numerical values of cohesion or bulk density.

\end{description}

\subsection{Database Architecture}
\begin{figure}
\centering
\includegraphics[width=0.97\linewidth]{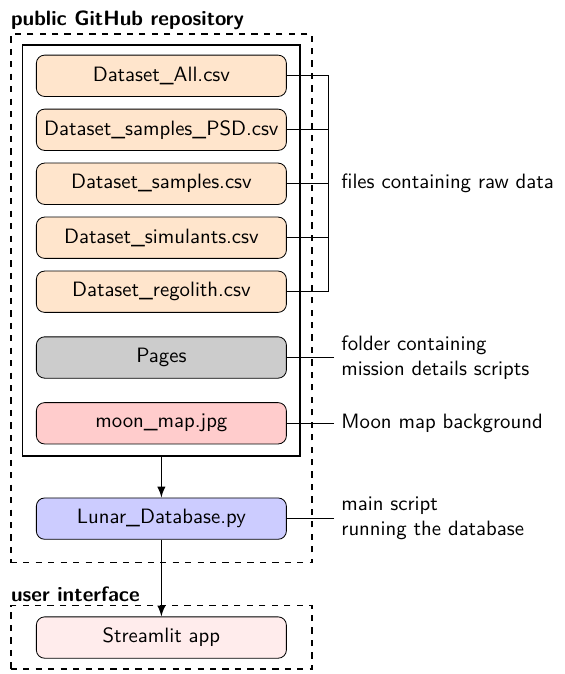}
\caption{\label{fig:DatabaseArchitecture}
\textbf{Schematic of the database architecture.}
Overview of the data files, processing scripts, and user interface forming the lunar regolith database.}
\end{figure}

The global architecture of the database is represented in Figure~\ref{fig:DatabaseArchitecture}.
It runs through a main python script that integrates all functionalities, including data filtering, visualization, and export options. 
The data is stored in \texttt{.csv} format in \texttt{Dataset\_Regolith} for true lunar regolith data, \texttt{Dataset\_Simulants} for simulants data,  \texttt{Dataset\_All} for data of lunar missions and simulants combined, and \texttt{Dataset\_Samples} and \texttt{Dataset\_Samples\_PSD} for data on the different lunar samples brought back to Earth and their particle size distribution. 
In the same folder, a \texttt{requirements.txt} file lists all necessary Python packages to run the database locally, a \texttt{README.md} file with installation and usage instructions, and a image \texttt{moon\_map.jpg} \cite{ernie_moonmap_2019} allowing for the lunar map visualization.
Finally, a folder titled \texttt{pages} contains the individual scripts for each mission reported in the database, providing more in-depth information on the context of data collection.

The database is accessible as a web application via a direct link (\url{https://lunar-regolith-database.streamlit.app/}) or by running a Python script developed with the \texttt{streamlit} package, which enables interactive data visualization through a web browser. 
The main script is located in a dedicated folder alongside supporting documents containing the raw datasets.

The database is organized into five main pages:

\begin{enumerate}
    \item \textbf{Lunar regolith data:}  
    This first page compiles all data collected from the various lunar missions. It includes:
    \begin{itemize}
        \item A searchable and sortable table containing metadata such as mission name, testing method, location, physical parameters, and reference source.
        \item An interactive plotting section with drop-down menus for selecting parameters and defining visualization modes (e.g., data ranges, minimum values, maximum values, or averages) with customizable legends.
        \item A filter panel (on the left side of the interface) that allows users to refine the displayed data according to specific criteria (e.g., mission, testing method, location, or parameter range).
        \item A lunar map displaying the geographic location on the Moon of each mission.
    \end{itemize}

    \item \textbf{Lunar regolith simulants data:}  
    Accessible via a button at the top of the filter panel, this page provides data on lunar regolith simulants and includes:
    \begin{itemize}
        \item A table compiling metadata for each simulant, such as developer, year of creation, simulant type (mare or highland), physical parameters, and reference source.
        \item An interactive plotting section and filter panel identical in functionality to those of the regolith data page, enabling comparative visualization.
    \end{itemize}

    \item \textbf{Lunar samples data:}  
    Accessible via a button at the top of the filter panel, this page provides data on the lunar samples brought back to Earth. It contains:
    \begin{itemize}
        \item A table compiling metadata for each sample catalogued. This includes the sample ID, weight, return container, and other parameters when available. 
        \item A table containing the particle size distribution of soil samples, when available.
        \item An interactive plotting section allowing to plot and compare the particle size distribution of different samples or generate a scatter plot comparing parameters such as sampling depth and median particle size.
    \end{itemize}

    \item \textbf{Mission details:}  
    The final page provides descriptive information on each mission included in the database. It features:
    \begin{itemize}
        \item A filter panel for selecting a specific mission.
        \item A section displaying general mission information, such as landing date, location, and primary scientific objectives.
        \item A table containing all the mechanical data on Lunar regolith obtained during the mission and sample analysis.
        \item A sample catalogue table including the sample ID, weight, return container, and other parameters when available.
    \end{itemize}

        \item \textbf{Combined dataset:}  
    This page, also accessible via the top navigation button, presents a unified table merging lunar regolith and simulant data. It allows:
    \begin{itemize}
        \item Direct comparison of both datasets through a single table containing mission, testing method, location, physical parameters, and reference information.
        \item Filtering capabilities to refine the dataset according to user-defined criteria.
    \end{itemize}

\end{enumerate}

All tables, filtered or not, can be exported directly from the interface in \texttt{.csv} format.

\section{Use-cases}\label{sec:usecases}

We illustrate typical applications of the database through three representative use-cases.
The tools implemented for filtering and visualization enables real-time updates of plots based on user-defined constraints, 
facilitating data exploration.
It also allows for rapid visual comparisons that can support model calibration or guide the interpretation of experimental results.
Here, data are filtered, then extracted directly from the database in \texttt{.csv} format 
and replotted,
as can be done by any ordinary user.

In the following, we always use the term \enquote{cohesion} as the intercept in Mohr-Coulomb, not as a direct measurement of interparticle cohesive forces.

\subsection{\textit{Use-case 1:} 
Comparing soil mechanics measured in-situ across lunar terrains}

\begin{figure*}[h!]
\centering
\includegraphics[width=\linewidth]{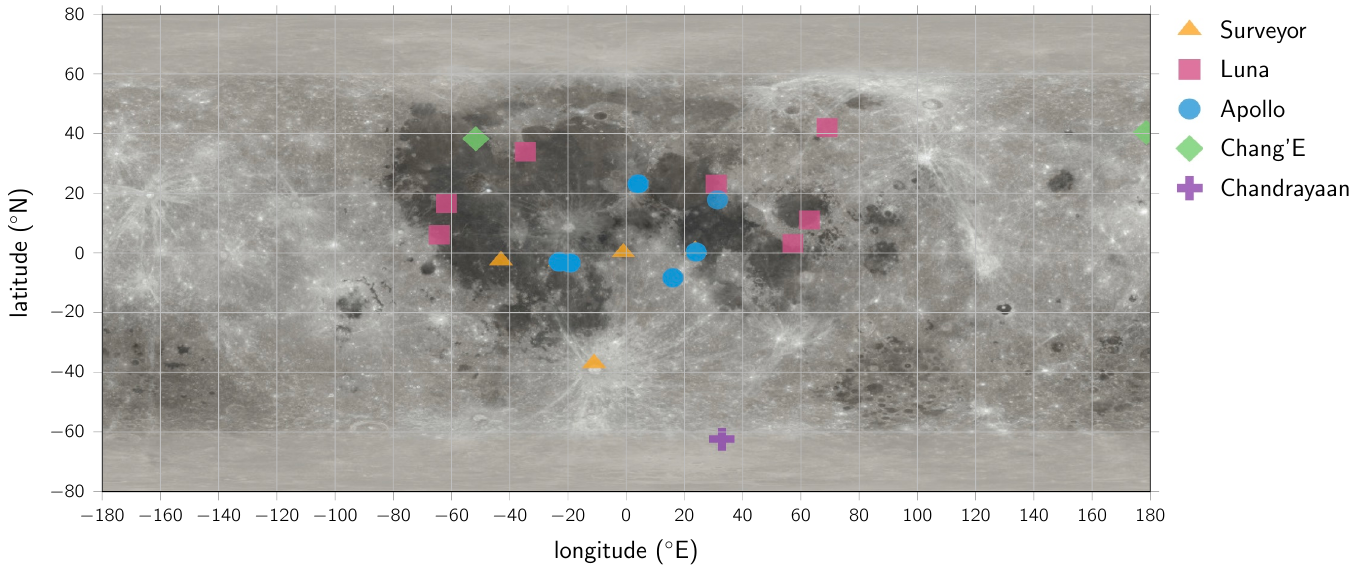}
\caption{\label{fig:MoonMapLandingSites}
\textbf{Moon map with missions landing sites.}
The global lunar albedo map shows the contrast between the dark lunar maria and the lighter highland regions. 
The cylindrical projection uses the selenographic coordinate system,
spanning from \SI{-180}{\degree}E to \SI{180}{\degree}E (left to right) 
and \SI{90}{\degree}N (North pole, top) to \SI{90}{\degree}S (South pole, bottom), 
centered on the mean sub-Earth point (\SI{0}{\degree}N, \SI{0}{\degree}E).
The map is reproduced with permission from NASA's Scientific Visualization Studio CGI Moon kit \cite{ernie_moonmap_2019}.
Each mission landing site is placed on the map (see legend for specific mission identification).
This map, including the mission landing sites, is directly extracted from the database website.
}
\end{figure*}

The lunar surface is broadly divided into two main terrain types: highlands and maria. 
The maria are relatively flat with low albedo, while the highlands are heavily cratered and mountainous with significantly higher albedo. 
Their formation processes explain their different topography, 
but also different chemical compositions: the highlands are dominated by feldspathic rocks, 
whereas the maria consist mainly of basaltic lavas rich in pyroxene \cite{LunarSourcebookCh2_Vaniman1991}. 
Besides, 
localized lunar pyroclastic deposits, 
produced by explosive volcanic eruptions,
are widespread across the Moon, occurring in both mare and highland regions \cite{Trang2022}.
Figure~\ref{fig:MoonMapLandingSites} presents a global Moon map, 
illustrating the contrast between the dark maria and the lighter highland regions,
including mission landing sites.
This map is extracted directly from the database.

\begin{figure}[h!]
\centering
\includegraphics[width=\columnwidth]{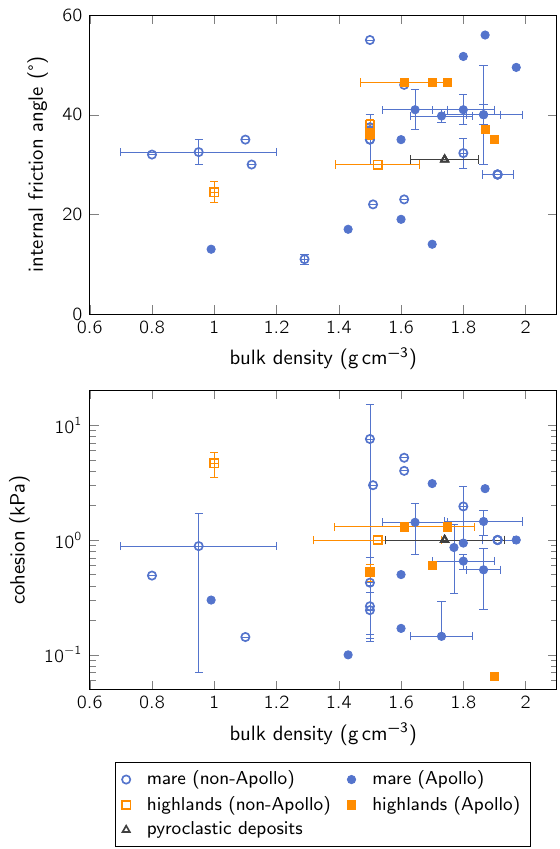}
\caption{\label{fig:ComparisonTerrains}
\textbf{Geomechanical properties of lunar regolith depending on type of lunar terrain.}
Internal friction angle (top) and cohesion (bottom), both parameters of the Mohr-Coulomb failure model, 
are shown as functions of bulk density. 
Cohesion values are plotted on a logarithmic scale ($y$-axis) to capture their wide range, 
spanning two orders of magnitude, from \SI{100}{\pascal} to \SI{10}{\kilo\pascal}. 
Filled symbols indicate data obtained during the Apollo missions, while open symbols represent data from all other missions.
}
\end{figure}

This distinction in geological evolution is also reflected in different regolith physical properties. 
Using the database, we investigate whether these different physical properties result in distinct mechanical responses. We generate two datasets, based only on in-situ measurements, 
%(from the Apollo and Luna missions, 
but collected by any testing method:
one with only measurements from mare sites and another from only highland sites. 
In Figure~\ref{fig:ComparisonTerrains}, 
we generate side-by-side visualizations of internal friction angle and cohesion (both Mohr-Coulomb model parameters), 
both plotted as a function of bulk density, 
for the two terrain types.

\begin{figure}[h!]
\centering
\includegraphics[width=\columnwidth]{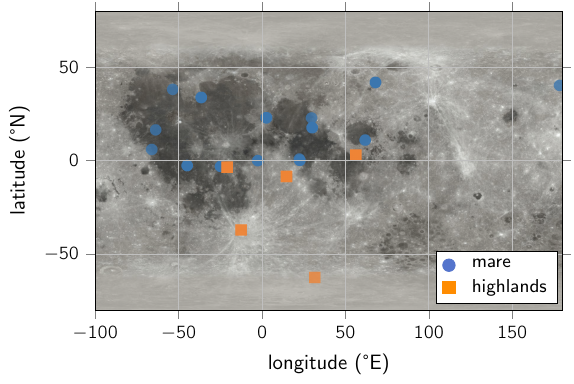}
\caption{\label{fig:MapCompTerrains}
\textbf{Moon map showing the geographic locations of missions as a function of lunar terrain.}
Locations are classified as mare (blue disks) and highland (orange squares);
the points highlight the spatial spread of the location of data collection.
}
\end{figure}

The Apollo missions program provided the first opportunity to systematically investigate these terrain types through sampling and in-situ experiments. 
The Apollo 11, 12, 15, and 17 missions landed in mare regions, while the Apollo 14 and 16 missions targeted highland terrains.
Figure~\ref{fig:MapCompTerrains} shows spatial spread of the location of data collection (note that it includes non-Apollo missions).

The Apollo 14 mission was the first of the program to explore lunar highland terrains, landing in the Fra Mauro formation. The properties of the soil during this mission were directly compared to those of the Apollo 11 and 12 missions, which had landed in mare regions. The results of the trench experiment \cite{herbert_apollo_1971} indicated that the highland regolith exhibited a much lower cohesion than that of the mare sites. The bulk density of the highland soil was also found to be lower than that of the mare sites, with a value of approximately \SI{1.5}{\g\per\cubic\cm} for the highlands, compared to \SI{1.75}{\g\per\cubic\cm} for the mare sites.

The Apollo 15 mission landed in a mare region and reported a bulk density of approximately \SI{2.0}{\g\per\cubic\cm}, higher than that of the Apollo 14 highland site \cite{calio_apollo_1971}. The values of the angle of internal friction for the Apollo 15 mare regolith ranged from 41 to 52 degrees, compared to approximately 35 degrees for the Apollo 14 highland site \cite{calio_apollo_1971}. The Apollo 16 mission, which landed in the Descartes highlands, provided further data on highland regolith properties. The bulk density measured at this site was approximately 1.7 $\mathrm{g/cm^3}$, displaying no significant difference with mare values. The angle of internal friction for the Apollo 16 highland regolith was around 46 degrees, again more similar to mare measurements than to highland regolith properties \cite{hinners_apollo_1972}. The last mission of the program, Apollo 17, landed in the Taurus-Littrow valley, a region characterized by both mare and highland features. The bulk density measured at this site was approximately \SI{1.6}{\g\per\cubic\cm}, while the angle of internal friction was found to be around 35 degrees \cite{hinners_apollo_1973}. These values were more consistent with mare regolith properties than with those of highland terrains.

Overall, 
the data from the Apollo missions reveal a large diversity among different results and locations on the Moon, 
even within each terrain type. 
It suggests that while some differences in regolith properties exist between mare and highland terrains, these differences may not be as pronounced as initially thought. 
The angle of internal friction, cohesion and  bulk density values for highland sites (Apollo 14 and 16) were found to be closer to those of mare sites (Apollo 11, 12, 15, and 17) than previously reported in earlier missions such as Surveyor VII. This finding indicates that local geological processes and impact history may play a more significant role in determining regolith properties than the broad classification of terrain type alone.
Based on the accessible in-situ measurements, 
the type of lunar terrain does not appear to induce strong differences in geomechanical soil properties, at least with respect to the parameters commonly used to characterize lunar soil.

A similar conclusion was drawn from observations of boulder tracks on the lunar surface: a qualitative comparison of track morphology across mare, highland, and pyroclastic deposits revealed no significant differences in their appearance, suggesting comparable geomechanical properties of the underlying materials \cite{Bickel2019}.

In conclusion,
we do not observe a clear distinction,
either because the intrinsic variability of lunar soil is larger than the differences between specific terrain types, or because the testing conditions and environments differ so much that these variations overshadow any real geological distinctions. The lack of repeated measurements on the same location and with the same methods ultimately prevents us from identifying the true origin of the discrepancies.

In contrast, soil bulk density emerges as a characteristic parameter
influencing the internal friction angle of lunar mare regolith.
This is statistically supported by a Spearman rank correlation analysis, selected to account for the limited sample size and presence of outliers, which demonstrates a strong positive relationship between bulk density and internal friction ($\rho = 0.759$, $p < 0.001$). 
Conversely, correlation analyses for cohesion and highland terrains did not yield statistically significant results, likely due to the inherent regolith variability and the limited amount of in-situ data currently available.
The density and mechanical strength of lunar regolith are known to vary with depth;
understanding this variation based on existing in-situ data is essential for predicting load-bearing capacity, excavation resistance, and stability during surface operations.

\subsection{\textit{Use-case 2:} 
Comparing in-situ data to returned samples measured on-ground}

The lunar environment differs profoundly from that we know on Earth: 
its surface gravity is only one-sixth of Earth's, it is surrounded by a thin surface-bound exosphere close to vacuum, and it undergoes regular micrometeoroid impacts in the absence of an atmosphere; 
it also lacks active weathering processes. 
These factors shape the physical and mechanical behavior of lunar regolith in ways that have no terrestrial analogue.
But beyond how these conditions modify the material itself, the testing environment may also affect measurement results.

The influence of gravity on granular behavior is now well established \cite{Murdoch2013,DAngelo2022,DAngelo2025}.
Here, we adopt a broader perspective: 
using data from previous missions, we assess whether geomechanical tests conducted in-situ on the lunar surface yield systematically different results from tests performed on returned samples.

\begin{figure}
\centering
\includegraphics[width=\linewidth]{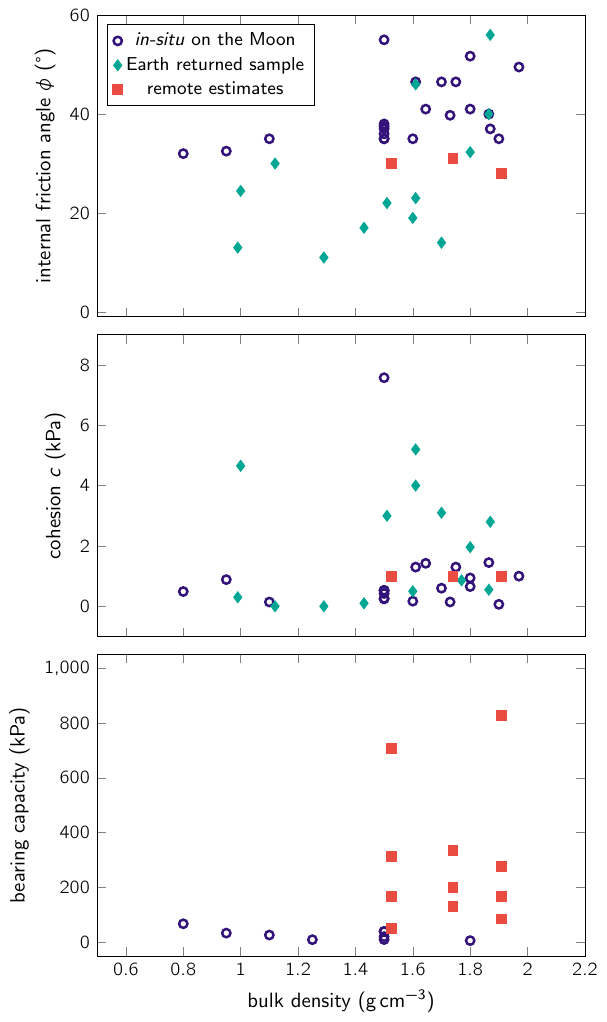}
\caption{\label{fig:EarthVsMoonTesting}
\textbf{Comparison of geomechanical properties measured on Earth versus the Moon.}
We compare three typical geomechanical properties (internal friction angle, cohesion, bearing capacity), as a function of bulk density, for true regolith tested in-situ, returned samples tested on ground, and properties estimated remotely. 
}
\end{figure}

In Figure~\ref{fig:EarthVsMoonTesting}, we pool results from all missions, ignoring differences in test type and sampling location on the Moon. Although the database makes these finer-grained analyses readily accessible, our goal here is more global: to evaluate whether lunar measurements diverge systematically from Earth-based geomechanical properties measurements.
We also include the dataset based on remote observations of boulder tracks \cite{Bickel2019},
which estimates cohesion at a fixed \SI{1}{\kilo\pascal}.

We see a clear pattern emerge: 
in-situ lunar data exhibits higher internal friction angles 
and often lower cohesion compared to measurements performed on returned samples. 
While the internal friction angle generally increases with bulk density, 
the relationship between cohesion and bulk density appears far less systematic.

Regarding bearing capacity, values estimated from boulder-track observations are up to one order of magnitude larger than those from experiments conducted directly on the lunar surface. 
This discrepancy must be attributed to differences in the tested material properties or the distinct measurement methods employed.

Finally, across all parameters considered, on-Earth measurements display substantially broader variability, 
underlining the strong influence of environmental conditions. This is a key issue in granular physics (and soft matter more broadly): because such materials are highly sensitive to gravity, what we often regard as an intrinsic property of the material can be, in practice, a characterization of the material \emph{within its environment}.
While this might not be inherently problematic, it becomes crucial to keep in mind when planning missions to markedly different environments,
as material properties may scale in fundamentally different ways from one environment to another.

\subsection{\textit{Use Case 3:} Selecting an appropriate lunar simulant}

Lunar regolith simulants have been developed for decades to reproduce the behavior of lunar soil for laboratory and engineering studies. 
Many lunar regolith simulants focus primarily on reproducing one or a few properties of true lunar regolith, for example its chemical composition, optical properties or particle size distribution. 
Depending on the specific process tested, it can be either perfectly sufficient or entirely inadequate. 

For example, challenges during in-situ experiments might be traced back to using simulants with inappropriate mechanical properties. 
During Apollo 15, astronauts encountered unexpected struggles with drill penetration and deep core sample retrieval, due to \enquote{unexpectedly hard material approximately \SI{1}{\meter} below the surface} \cite{Ap15MissionReport1971}.
The drill stems did not penetrate at expected rates, and underwent mechanical dysfunction  due to torques that were much higher than expected following testing on-ground, in terrestrial soil. Concretely, extracting the deep core sample required both crewmen lifting on the drill handles,  costing far more time and effort than planned, and ultimately limiting the crew's ability to explore other Moon locations \cite{Ap15MissionReport1971}.

This example illustrates the importance of 
having knowledge about the mechanical properties and behavior of the lunar surface
when designing a mission.
To test tools or scientific instruments to be used on the Moon, it is often particularly useful to know which lunar simulant is most representative of a particular region of the Moon, in flow or mechanical properties.
Appropriate lunar simulants can also help to improve the planning of operations and to interpret acquired data.

The database includes data for existing simulants, enabling direct comparisons with either lunar regolith or other available simulants. 
Due to the large number of existing simulants, as well as the rate at which new simulants are produced and (partially) characterized,  
it is challenging to maintain a comprehensive and up-to-date catalogue referencing all available simulants.
We focus on widely-used simulants whose characterization has been published in peer-reviewed articles,
and update the list based on individual requests, provided that the characterization data has been previously published in a peer-reviewed article. 
We also include original data obtained from direct shear testing of the European lunar regolith simulant EAC-1A \cite{Gasteiner2025}, directly demonstrating how the database can be used to compare properties of a new material with in-situ lunar data and other simulants.

\begin{figure}
\centering
\includegraphics[width=\columnwidth]{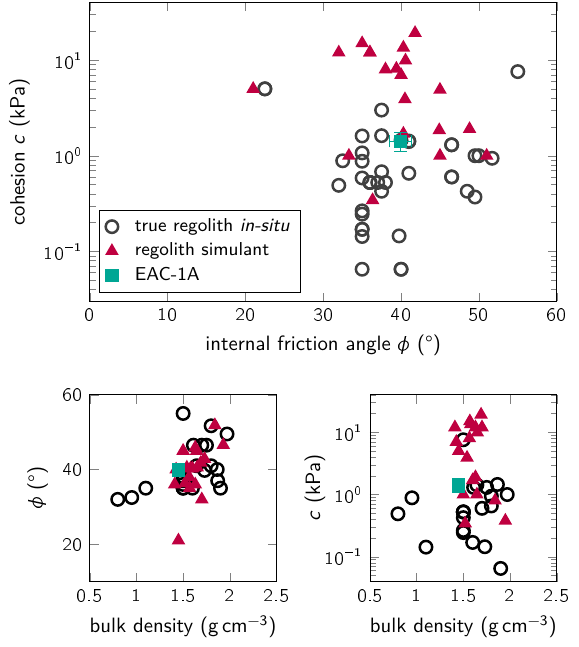}
\caption{\label{fig:RegolithVsSimulants}
\textbf{Comparison of true regolith in-situ to regolith simulants, tested on Earth.}
The Mohr-Coulomb parameters (cohesion as a function of angle of internal friction) are given in the main panel,
for true lunar regolith and lunar simulants;
below, the dependence of each parameter is given as a function of the material's bulk density.
}
\end{figure}

In Figure~\ref{fig:RegolithVsSimulants}, we see that the cloud of points corresponding to true lunar regolith tested on the Moon
is clearly below that of regolith simulants,
showing materials with lower cohesion, albeit relatively similar angle of internal friction. 
This observation is supported by a comparative analysis of the two groups. Given the non-normal distribution and unequal variances of the datasets, a Welch t-test is used. 
This confirms that the cohesion of in-situ regolith is significantly lower than that of the simulants ($p < 0.001$),
beyond
%This result indicates that the observed discrepancy is a systemic difference rather than a product of 
random sampling error.

Still focusing specifically on cohesion,
we see however that a lower cohesion in the Mohr-Coulomb sense does not seem to be directly correlated with a lower bulk density.

A possible explanation lies in the particles' morphology: lunar regolith particles tend to be highly angular and geometrically complex, leaving more void space between particle surfaces.
This reduces the number of true contact points, thereby weakening surface-dominated cohesive forces, even at equivalent bulk densities.

For the friction angle analysis, considering the non-normality but equality of variances of the two datasets, a Welch t-test was used for the comparison. It yielded non-significant results ($p=0.850$), indicating that the angle of internal friction of the simulants is statistically indistinguishable from that of the in-situ regolith. 
This suggests that while existing lunar regolith simulants struggle to replicate the cohesion of the lunar soil, they successfully replicate the frictional characteristics of the material.

When comparing our results for the EAC-1A simulant, we observe that, like other simulants, it successfully mimics the friction angle and bulk density of true lunar soil in-situ, but tends to exhibit higher cohesion, although lower than that of most other simulants.

This comparison highlights an important limitation of terrestrial regolith simulants. 
Because their formation processes differ fundamentally from those of lunar materials, 
simulants diverge in certain properties. It is therefore essential to identify and reproduce the \emph{relevant} physical parameters as faithfully as possible; 
otherwise, mission preparation performed with an inappropriate simulant risks leading to unexpected failures on planetary surfaces.

\section{Discussion}\label{sec:discussion}

\subsection{Data Coverage and Tool Performance}
The current version of the Lunar Regolith Database includes over 200 entries spanning all six Apollo landing sites, several Luna mission sites, and selected robotic mission data. The dataset covers bulk density values ranging from \SIrange{0.7}{2.3}{\g\per\cubic\cm}, angles of internal friction from \SIrange{20}{55}{\degree}, and cohesion from negligible up to approximately \SI{5}{\kilo\pascal}, depending on sampling depth and location. 
Coverage varies by parameter and mission: for example, most bulk density measurements come from Apollo core samples, while cohesion and internal friction are more frequently derived from penetrometer and trench experiments.
The database also includes over 2000 entries on lunar samples brought back to Earth, including all samples from Apollo missions and some from Luna missions, detailing the type, return container, and grain size distribution when available.
The interactive filtering and visualization tool allows users to explore this dataset efficiently, generating plots and maps in real time based on selected constraints. For instance, researchers can quickly compare regolith properties across Mare and Highland terrains, examine depth-dependent trends, or identify suitable simulants for laboratory experiments. The platform thus provides a critical bridge between historical mission data and contemporary analytical and engineering needs.

\subsection{Limitations and Uncertainties}
While the database significantly improves accessibility and standardization of lunar regolith data, several limitations remain. The precision and accuracy of the original measurements vary widely due to differences in mission instrumentation, sampling methods, and experimental conditions. Some values, particularly those derived from touchdown analyses or optical observations, are indirect estimates rather than direct measurements. Future work will include uncertainty quantification and expansion to incorporate remote-sensing derived properties and analog simulant data for laboratory validation.

Additionally, the database currently focuses on mechanical properties measured in-situ or from returned samples. Expanding the database to include additional information on returned samples experiments and results, chemical composition and other properties would further enhance its utility for scientific and engineering applications.

\subsection{Future Work}
The database is intended to evolve in time. Future updates will include data from upcoming missions such as Artemis and Chang'E 6--7, additional laboratory analyses of regolith simulants, and remote-sensing derived properties that can fill existing spatial gaps. We also envision integrating this tool with planetary GIS platforms and simulation environments for \gls{isru} systems, lunar base planning, and regolith processing technologies.

Community contributions are actively encouraged, provided that submitted data meet established quality and documentation standards.

\section{Conclusion}\label{sec:conclusion}
We present a consolidated and accessible database of lunar regolith properties, paired with an interactive interface for data exploration and visualization. By centralizing decades of mission data and enabling rapid filtering, mapping, and comparison, the database supports both retrospective scientific analysis and forward-looking mission design.

The use-cases presented 
illustrate practical applications of the tool for research and engineering. Beyond facilitating fundamental studies in planetary science, this database has direct relevance for rover mobility analysis, lander design, \gls{isru} system planning, and lunar construction activities.

As an open-access resource, the Lunar Regolith Database is designed to evolve with new mission data, laboratory studies, and community contributions, 
providing a growing foundation for safe and efficient exploration of the Moon.

\bmsection*{Acknowledgments}

N.\ M.\ and L.\ G.\ acknowledge funding from the European Research Council (ERC) 
GRAVITE project (Grant Agreement \textnumero ~1087060) and financial support from 
the French National Centre for Space Studies (CNES) in the context of the HERA space mission.  

O.\ D'A.\ acknowledges financial support from 
CNES under the CNES fellowship 24-357
and APR \textnumero~10678 (2025).

This publication is part of the project \textit{Influence of gravity on granular flows} 
with file \textnumero 21898 of the research programme VENI 
which is (partly) financed by the Dutch Research Council (NWO) 
under the grant
\href{https://doi.org/10.61686/DBYVT15219}.

%\bmsection*{Financial disclosure}
%None reported.
%
%\bmsection*{Conflict of interest}
%The authors declare no potential conflict of interests.

%\bibliographystyle{plainnat}   
\bibliography{wileyNJD-APS.bib}

%\bmsection*{Supporting information}
%
%Additional supporting information may be found in the
%online version of the article at the publisher's website.
%

%\appendix

%\nocite{*}% Show all bib entries - both cited and uncited; comment this line to view only cited bib entries;

%\bmsection*{Author Biography}

%\begin{biography}{\includegraphics[width=76pt,height=76pt,draft]{empty}}{
%{\textbf{Author Name.} Please check with the journal's author guidelines whether
%author biographies are required. They are usually only included for
%review-type articles, and typically require photos and brief
%biographies for each author.}}
%\end{biography}

\end{document}